\documentclass{article}
\usepackage{amssymb, amsmath}
\numberwithin{equation}{section}
\newtheorem{theorem}{Theorem}[section]
\newtheorem{definition}{Definition}[section]
\newtheorem{remark}{Remark}[section]
\newtheorem{proposition}{Proposition}[section]
\newtheorem{corollary}{Corollary}[section]
\newtheorem{lemma}{Lemma}[section]
\def\div{{\rm div}}
\def\supp{{\rm supp}}
\def\Log{{\rm Log}}
\def\max{{\rm max}}
\def\measure{{\rm measure}}
\def\det{{\rm det}}
\begin{document}
\title{The Einstein-Vlasov-Maxwell(EVM) System with Spherical Symmetry}
\author{P. Noundjeu \\
Department of Mathematics, Faculty of Science, University of
Yaounde 1,\\ PO Box 812, Yaounde, Cameroun \\ {e-mail:
noundjeu@uycdc.uninet.cm or pnoundjeu@yahoo.fr}}
\date{}
\maketitle
\begin{abstract}
We look for the global in time solution of the Cauchy problem
corresponding to the asymptotically flat spherically symmetric
E.V.M system with small initial data. Using an estimate, we also
prove that if solution of the system stated above develops a
singularity at all time, then the first one has to appear at the
center of symmetry.
\end{abstract}
\section{Introduction}
The EVM system models the time evolution of self-gravitating
collisionless charged particles in the context of general
relativity. The particle could be for instance the electrons in a
plasma. As it is proved in \cite{noundjeu3}, that system in the
context of spherical symmetry has nice properties. Firstly, the
electromagnetic field $F_{\alpha \beta}$ created by the fast
moving particles reduces to its electric part and secondly if $f =
0$ then $F_{\alpha \beta} = 0$ and the latter is not true in the
case without spherical symmetry, where it is possible to have non
trivial source-free solutions of the Maxwell equations.

In \cite{noundjeu2}, the authors established with the help of an
iterative scheme one local existence theorem and a result on the
continuation criterion of solutions is deduced. Note that these
results are valid only for the Cauchy problem with small initial
data. Here, we aim to establish a global existence theorem of
solutions for the asymptotically flat spherically symmetric EVM
system. To do so, contrary to the uncharged case, we defined a set
of initial data in such a way that for fixed $\overset{\circ}{f}$,
the solution $(\overset{\circ}{\lambda}, \overset{\circ}{e})$ of
the constraint equations(see \cite{noundjeu1}) satisfies the
condition that $\overset{\circ}{\lambda}$ is bounded in the
$L^{\infty}$-norm. The latter is possible, provide that the charge
$q$ is small. Note that in the above, $\overset{\circ}{\lambda}$
(resp. $\overset{\circ}{e}$, $\overset{\circ}{f}$) is the initial
datum of the metric function $\lambda(t, r)$(resp. electric field
$e(t, r )$, distribution function of the particles $f$). So, the
global existence theorem we obtain, generalize that established by
G. Rein and AD. Rendall in \cite{rein3}.

Next, for the related literature, global existence for the
Vlasov-Maxwell system without symmetry is not known in general.
The most general result which has been proved is for a case with
one-dimensional symmetry group \cite{glassey}. Global existence
for the Yang-Mills equations in Minkowski space was proved by
Eardley and Moncrief \cite{eardley} and another proof by quite
different methods was given by Klainerman and Machedon
\cite{klainerman}. Global existence for the Yang-Mills equations
on a general globally hyperbolic spacetime was proved by
Chru\'sciel and Shatah \cite{chrusciel}.

Now, in the case of spherical symmetry, the initial boundary value
problem for the asymptotically flat EVM system admits the global
solution and the obtained spacetime is geodesically complete i.e
each geodesic is defined on $\mathbb{R}$. So, it is well known in
general relativity that this kind of solution does not develop a
singularity. Concerning the generic initial data, it is proved by
A.D. Rendall \cite{rendall2} that a singularity may occur. Also,
there is a related result of G. Rein , A.D Rendall and J.A
Schaeffer \cite{rein2} which states that for the asymptotically
flat spherically symmetric Einstein-Vlasov system, if solutions
develop a singularity at all time then the first one has to occur
at the center of symmetry, and also if particles remain away from
the center, the local solution of Cauchy problem for that system
with a given initial data can be extended to obtain the global
one. That results concern uncharged particles.

It would be interesting to see what happens in the case of charged
particles. More precisely, is it possible to generalize the above
results in the context of charged particles? The answer is yes and
this is the second important result of this paper, the first one
being the global existence theorem we stated earlier.

The paper proceeds as follows. In Sect.2, we formulate the Cauchy
problem. In Sect.3, we show that if the charge is small then one
can choose a set of initial data so that solutions depend
continuously on the latter. In Sect.4, we establish a global
existence theorem of solution for the corresponding initial
boundary value problem and the obtained spacetime is geodesically
complete. In Sect.5, we establish the main estimates we use to
prove the global existence theorem in the case that all particles
remain away from the center. In Sect.6, we prove a local existence
theorem and a continuation criterion for the exterior problem.
Sect.7 contains the essential result obtained using all what have
been established in the last two sections above.
\section{The Cauchy Problem}
We consider fast moving collisionless particles with charge $q$.
The basic spacetime is $(\mathbb{R}^{4}, g)$, with $g$ a
Lorentzian metric with signature $(-, +, +, +)$. In what follows,
we assume that Greek indices run from $0$ to $3$ and Latin indices
from $1$ to $3$, unless otherwise specified. We also adopt the
Einstein summation convention, both the gravitational constant and
the speed of light are setting to unity. Now, Using the assumption
of spherical symmetry, one obtains with the help of the Killing
equation for instance that $g$ can be taken of the following form
\cite{papapetrou}:
\begin{equation} \label{eq:2.1}
ds^{2} = - e^{2\mu}dt^{2} + e^{2\lambda}dr^{2} + r^{2}(d\theta^{2}
+ (\sin \theta)^{2}d\varphi^{2})
\end{equation}
where $\mu = \mu(t, r)$; $\lambda = \lambda(t, r)$; $t \in
\mathbb{R}$; $r \in [0, + \infty[$; $\theta \in [0, \pi]$;
$\varphi \in [0, 2\pi]$. One shows (see \cite{noundjeu3}) that the
spherically symmetric EVM  system can be written as the following
P.D.E system in $\lambda$, $\mu$, $f$, $e$:
\begin{equation} \label{eq:2.2}
e^{-2\lambda}(2r\lambda' - 1) + 1 = 8\pi r^{2}\rho
\end{equation}
\begin{equation} \label{eq:2.3}
e^{-2\lambda}(2r\mu' - 1) + 1 = 8\pi r^{2}p
\end{equation}
\begin{equation} \label{eq:2.4}
\frac{\partial f}{\partial t} + e^{\mu - \lambda} \frac{v}{\sqrt{1
+ v^{2}}} . \frac{\partial f}{\partial \tilde{x}} - \left( e^{\mu
- \lambda} \mu' \sqrt{1 + v^{2}} + \dot{\lambda}
\frac{\tilde{x}.v}{r} - q e^{\lambda + \mu} e(t, r)
\right)\frac{\tilde{x}}{r}.\frac{\partial f}{\partial v} = 0
\end{equation}
\begin{equation} \label{eq:2.5}
\frac{\partial}{\partial r}(r^{2}e^{\lambda}e) =
qr^{2}e^{\lambda}M
\end{equation}
where $\lambda' = \frac{d\lambda}{dr}$; \, $\dot{\lambda} =
\frac{d\lambda}{dt}$ and:
\begin{equation} \label{eq:2.6}
\rho(t, \tilde{x}) = \int_{\mathbb{R}^{3}} f(t,
\tilde{x},v)\sqrt{1 + v^{2}} dv + \frac{1}{2} e^{2\lambda(t,
\tilde{x})} e^{2}(t, \tilde{x})
\end{equation}
\begin{equation} \label{eq:2.7}
p(t,\tilde{x}) = \int_{\mathbb{R}^{3}} \left(
\frac{\tilde{x}.v}{r} \right)^{2} f(t, \tilde{x}, v)
\frac{dv}{\sqrt{1 + v^{2}}} - \frac{1}{2} e^{2\lambda(t,
\tilde{x})} e^{2}(t, \tilde{x})
\end{equation}
\begin{equation} \label{eq:2.8}
M(t, \tilde{x}) = \int_{\mathbb{R}^{3}} f(t, \tilde{x}, v)dv.
\end{equation}
Here (\ref{eq:2.2}) and (\ref{eq:2.3}) are the Einstein equations
for $\lambda$ and $\mu$, (\ref{eq:2.4}) is the Vlasov equation for
$f$, $f$ being the distribution function of the charged particles
which is defined on the mass-shell and (\ref{eq:2.5}) is the
Maxwell equation for $e$. Here $\tilde{x} := (x^{i})$ and $v$
belong to $\mathbb{R}^{3}$, $r := \mid \tilde{x} \mid$,
$\tilde{x}.v$ denotes the usual scalar product of vectors in
$\mathbb{R}^{3}$, and $v^{2} := v.v$. The distribution function
$f$ is assumed to be invariant under simultaneous rotations of
$\tilde{x}$ and $v$, hence $\rho$, $p$ and $M$ can be regarded as
functions of $t$ and $r$. It is assumed that $f(t)$ has compact
support for each fixed $t$. We are interested in asymptotically
flat spacetime with a regular center, which leads to the boundary
conditions that:
\begin{equation} \label{eq:2.9}
\lim_{r \to \infty}\lambda(t, r) = \lim_{r \to \infty}\mu(t, r) =
\lambda(t, 0) = \lim_{r \to \infty}e(t, r) = e(t, 0) = 0
\end{equation}
Now, define the initial data by:
\begin{equation} \label{eq:2.10}
\begin{cases}
f(0, \tilde{x}, v) = \overset{\circ}{f}(\tilde{x}, v); \quad
\lambda(0, \tilde{x}) =\overset{\circ}{\lambda}(\tilde{x}) =
\overset{\circ}{\lambda}(r)\\
\mu(0, \tilde{x}) = \overset{\circ}{\mu}(\tilde{x}) =
\overset{\circ}{\mu}(r); \quad e(0, \tilde{x}) =
\overset{\circ}{e}(\tilde{x}) = \overset{\circ}{e}(r)
\end{cases}
\end{equation}
with $\overset{\circ}{f}$ being a $C^{\infty}$ function with
compact support, which is nonnegative and spherically symmetric,
i.e
\begin{displaymath}
\forall A \in SO(3), \, \forall (\tilde{x}, v) \in \mathbb{R}^{6},
\, \overset{\circ}{f}(A\tilde{x}, A v) =
\overset{\circ}{f}(\tilde{x}, v).
\end{displaymath}
We aim to solve the initial boundary value problem (\ref{eq:2.2}),
(\ref{eq:2.3}), (\ref{eq:2.4}), (\ref{eq:2.5}), (\ref{eq:2.9}) and
(\ref{eq:2.10}).
\begin{remark} \label{R:2.1}
Note that the assumption $e(t, 0) = 0$ makes sense. To see the
latter, we use the fact that the electric field $E(t, \tilde{x})
:= e(t, \tilde{x})\frac{\tilde{x}}{\mid \tilde{x}\mid}$ is
spherically symmetric in the $\tilde{x}$ variable i.e $AE = E$ for
every $A \in SO(3)$ and once \\ $E(t, 0) = 0$, one easily deduces
the announced result.
\end{remark}
\section{Continuous dependence on the initial data}
In this section, we prove that solutions depend continuously on
their initial data. Besides being of interest in itself for
physically viable theory ( cf.[\cite{wald}, p. 243f]), the results
of this section will be applied in the proof of global existence
for small data in the next section. Before doing so, we recall the
local existence theorem we proved in \cite{noundjeu2} on which our
present result relies. The constraint equations are obtained
setting $t = 0$ in (\ref{eq:2.2}), (\ref{eq:2.3}) and
(\ref{eq:2.5}). Using perturbation techniques, it is proved in
\cite{noundjeu1} that for $\overset{\circ}{f}$ small the
constraint equations has a global smooth (i.e $C^{\infty}$)
solution. First of all, we need to make precise the notion of
regularity we use in the sequel.
\begin{definition}
A solution $(f, \lambda, \mu, e)$ of (\ref{eq:2.2}),
(\ref{eq:2.3}), (\ref{eq:2.4}) and (\ref{eq:2.5}),
 is said to be regular if:
\begin{itemize}
\item[(i)] $f$ is nonnegative, spherically symmetric, $C^{1}$, and
$f(t)$ is compactly supported for all $t$,
\item[(ii)] $\lambda \geq 0$, $\mu \leq 0$, and $\lambda$, $\mu$,
$\lambda'$, $\mu'$ and $e$ are $C^{1}$ which $\lambda$, $\mu$ and
$e$ satisfying the boundary condition (\ref{eq:2.9}).
\end{itemize}
\end{definition}
\begin{theorem}[local existence, continuation criterion] \label{T:3.1}
Let $\overset{\circ}{f} \in C^{\infty}(\mathbb{R}^{6})$ be
nonnegative, compactly supported and spherically symmetric such
that:
\begin{equation} \label{eq:3.1}
8\pi \int_{0}^{r} \left( \int_{\mathbb{R}^{3}}
\overset{\circ}{f}(r, v)\sqrt{1 + v^{2}}dv \right)ds < r
\end{equation}
Let $\overset{\circ}{\lambda}, \overset{\circ}{\mu},
\overset{\circ}{e} \in C^{\infty}(\mathbb{R}^{3})$ be a regular
solution of the constraint equations. Then there exists a unique
regular solution $(\lambda, \mu, f, e)$ of the asymptotically flat
spherically symmetric EVM system with initial data
$(\overset{\circ}{\lambda}, \overset{\circ}{\mu},
\overset{\circ}{f}, \overset{\circ}{e})$ on a maximal interval $I
\subset \mathbb{R}$ of existence which contains $0$. If
\begin{equation*}
\sup \{ \mid v \mid |\, (t, \tilde{x}, v) \in \supp f, \quad t
\geq 0 \} < + \infty
\end{equation*}
then $\sup I = + \infty$, if
\begin{equation*}
\sup \{ \mid v \mid |\, (t, \tilde{x}, v) \in \supp f, \quad t
\leq 0 \} < + \infty
\end{equation*}
then $\inf I = - \infty$.
\end{theorem}
\textbf{Proof:} See \cite{noundjeu2}.

For $r_{0} > 0$, $u_{0} > 0$ and  $\Lambda > 0$, we consider the
following set of initial data:
\begin{align*}
&D := \{ (\overset{\circ}{f}, \overset{\circ}{\lambda},
\overset{\circ}{e}) \in C^{\infty}(\mathbb{R}^{6}) \times
(C^{1}([0, + \infty[))^{2}, \quad \overset{\circ}{f} \geq 0,
\text{spherically symmetric},\\
& \qquad \text{and satisfies} \quad (\ref{eq:3.1}) \quad \supp
\overset{\circ}{f} \subset B(r_{0}) \times B(u_{0}) \text{and}
\quad
(\overset{\circ}{\lambda},\overset{\circ}{e})\\
& \qquad \text{is a regular solution of the constraints with} \,
\parallel \overset{\circ}{\lambda} \parallel_{L^{\infty}} \leq \Lambda \}.
\end{align*}
Fix the solution $(\lambda_{g}, \mu_{g}, e_{g}, g)$ of
(\ref{eq:2.2}), (\ref{eq:2.3}), (\ref{eq:2.4}) and (\ref{eq:2.5})
with initial datum $(\overset{\circ}{\lambda}_{g},
\overset{\circ}{e}_{g}, \overset{\circ}{g}) \in D$ and right
maximal existence interval(r.m.e.i) $[0, T_{g}[$. We aim to
control the distance of another solution $(\lambda_{f}, \mu_{f},
e_{f}, f)$ from $(\lambda_{g}, \mu_{g}, e_{g}, g)$ and the
relation between the maximal existence times $T_{f}$ and $T_{g}$
in terms of distance of the initial data, $[0, T_{f}[$ being the
r.m.e.i of $(\lambda_{f}, \mu_{f}, e_{f}, f)$; the whole argument
would also work for $t < 0$. To do so in our context, we first
have to control the distance between two solutions
$(\overset{\circ}{\lambda}_{f}, \overset{\circ}{e}_{f})$,
$(\overset{\circ}{\lambda}_{g}, \overset{\circ}{e}_{g})$ of
constraint equations and the essential tool we use is the fact
that we can construct a set of initial data
$(\overset{\circ}{\lambda}, \overset{\circ}{e})$ such a way as the
$L^{\infty}$-norm of $\overset{\circ}{\lambda}$ is uniformly
bounded. This comes from the continuous dependence of solutions of
the constraint equations on parameter $q$, when $q$ is small as it
is shown in \cite{noundjeu1}.
\subsection{Continuous dependence of solutions for the
constraint equations} Let us give the main result of this section:
\begin{proposition} \label{P:3.1}
Consider $(\overset{\circ}{f}, \overset{\circ}{\lambda}_{f},
\overset{\circ}{e}_{f}), (\overset{\circ}{g},
\overset{\circ}{\lambda}_{g}, \overset{\circ}{e}_{g}) \in D$.
Given a sufficiently small real number $\varepsilon > 0$, if $d :=
\parallel \overset{\circ}{f} - \overset{\circ}{g}
\parallel_{L^{\infty}} < \varepsilon$, then
\begin{equation} \label{eq:3.2}
\parallel \overset{\circ}{\lambda}_{f}\parallel_{L^{\infty}},
\quad \parallel \overset{\circ}{e}_{f}\parallel_{L^{\infty}} \leq
C
\end{equation}
\begin{equation} \label{eq:3.3}
\begin{aligned}
&\parallel e^{\overset{\circ}{\lambda}_{f}} -
e^{\overset{\circ}{\lambda}_{g}} \parallel_{L^{\infty}}, \quad
\parallel e^{\overset{\circ}{\lambda}_{f}} \overset{\circ}{e}_{f}-
e^{\overset{\circ}{\lambda}_{g}} \overset{\circ}{e}_{g}
\parallel_{L^{\infty}}, \quad \parallel \overset{\circ}{e}_{f} -
\overset{\circ}{e}_{g} \parallel_{L^{\infty}}\\
&\parallel \overset{\circ}{\lambda}_{f} -
\overset{\circ}{\lambda}_{g} \parallel_{L^{\infty}}, \quad
\parallel e^{2\overset{\circ}{\lambda}_{f}} \overset{\circ}{e}_{f}^{2}-
e^{2\overset{\circ}{\lambda}_{g}} \overset{\circ}{e}_{g}^{2}
\parallel_{L^{\infty}}, \quad \parallel e^{2\overset{\circ}{\lambda}_{f}} -
e^{2\overset{\circ}{\lambda}_{g}} \parallel_{L^{\infty}} \leq Cd
\end{aligned}
\end{equation}
where the constant $C$ depends on $r_{0}, u_{0},
\overset{\circ}{g}, \Lambda$ and not on $\overset{\circ}{f}$.
\end{proposition}
\textbf{Proof :} Take $(\overset{\circ}{f},
\overset{\circ}{\lambda}_{f}, \overset{\circ}{e}_{f}) \in D$, with
$d < \varepsilon$. The bound of $\overset{\circ}{\lambda}_{f}$
comes immediately from the definition of $D$. Next, setting $t =
0$ in (\ref{eq:2.5})and replacing $f$ by $\overset{\circ}{f}$, we
obtain by integration:
\begin{equation*}
\overset{\circ}{e}_{f}(r) =
\frac{q}{r^{2}}e^{-\overset{\circ}{\lambda}_{f}
(r)}\int_{0}^{r}s^{2}e^{\overset{\circ}{\lambda}_{f}(s)}
\overset{\circ}{M}_{f}(s)ds
\end{equation*}
where $\overset{\circ}{M}_{f}$ is defined as in (\ref{eq:2.8}),
replacing $f$ by $\overset{\circ}{f}$. Distinguishing the cases $r
\leq r_{0}$ and $r \geq r_{0}$, we obtain, since
$\overset{\circ}{f}$ is with compact support, $\mid
\overset{\circ}{e}(r) \mid  \leq C$, where $C = C(r_{0}, u_{0},
\overset{\circ}{g}, \Lambda)$ is a constant , and (\ref{eq:3.2})
holds. To prove inequalities (\ref{eq:3.3}), one writes:
\begin{equation*}
e^{\overset{\circ}{\lambda}_{f}} -
e^{\overset{\circ}{\lambda}_{g}} = \frac{2}{r}
e^{2\overset{\circ}{\lambda}_{g} + \overset{\circ}{\lambda}_{f}}
\frac{\overset{\circ}{m}_{f} - \overset{\circ}{m}_{g}}{1 +
e^{\overset{\circ}{\lambda}_{g} - \overset{\circ}{\lambda}_{f}}}
\end{equation*}
where $\overset{\circ}{m}_{f} = m_{f}(0, r)$, $m(t, r)$ being the
mass function:
\begin{equation} \label{eq:3.4}
m(t, r) := 4\pi \int_{0}^{r}s^{2}\rho(t, s)ds
\end{equation}
and obtain using the Gronwall lemma:
\begin{equation*}
\mid e^{\overset{\circ}{\lambda}_{f}} -
e^{\overset{\circ}{\lambda}_{g}} \mid(r) \leq C\parallel
\overset{\circ}{f} - \overset{\circ}{g}
\parallel_{L^{\infty}}\exp \left( C\int_{0}^{r}(r - s)^{2}\mid
\overset{\circ}{M}_{f}(s) \mid ds \right) \leq Cd
\end{equation*}
We also deduce, using the mean value theorem and for $\varepsilon$
sufficiently small:
\begin{align*}
\mid \overset{\circ}{\lambda}_{f} - \overset{\circ}{\lambda}_{g}
\mid(r) &= \mid \Log \, e^{\overset{\circ}{\lambda}_{f}} - \Log \,
e^{\overset{\circ}{\lambda}_{g}} \mid(r)\\
&\leq \sup\left\{ \frac{1}{z} | \mid z -
e^{\overset{\circ}{\lambda}_{g}} \mid(r) \leq C\varepsilon
\right\}\mid e^{\overset{\circ}{\lambda}_{f}}
 - e^{\overset{\circ}{\lambda}_{g}} \mid(r)\\
&\leq Cd,
\end{align*}
and the rest of inequalities in (\ref{eq:3.3}) follow immediately
from those above. Thus, this ends the proof of Proposition 3.1.
\subsection{Continuous dependence of solutions for the
Einstein-Vlasov-Maxwell on initial data} We now state the
essential result of this section:
\begin{theorem} \label{T:3.2}
There exists a constant $\varepsilon > 0$, a positive increasing
function $\xi \in C([0, T_{g}[)$, and a positive decreasing
function $\sigma \in C(]0, \varepsilon[)$ such that
\begin{equation} \label{eq:3.5}
\lim_{\beta \to 0}\sigma(\beta) = T_{g}
\end{equation}
and for any solution $(\lambda_{f}, \mu_{f}, e_{f}, f)$ with the
initial datum $(\overset{\circ}{f}, \overset{\circ}{\lambda}_{f},
\overset{\circ}{e}_{f}) \in D$ satisfying $d := \parallel
\overset{\circ}{f} - \overset{\circ}{g}
\parallel_{L^{\infty}} < \varepsilon$, we have the estimates:
\begin{equation} \label{eq:3.6}
T_{f} > \sigma(d)
\end{equation}
\begin{equation} \label{eq:3.7}
\begin{aligned}
&\parallel f(t) - g(t)\parallel_{L^{\infty}} + \parallel
\lambda_{f}(t) - \lambda_{g}(t) \parallel_{L^{\infty}} + \parallel
\mu_{f}(t) - \mu_{g}(t) \parallel_{L^{\infty}} + \parallel
e_{f}(t) - e_{g}(t) \parallel_{L^{\infty}}\\
& + \parallel e^{2\lambda_{f}(t)} - e^{2\lambda_{g}(t)}
\parallel_{L^{\infty}} + \parallel \dot{\lambda}_{f}(t) -
\dot{\lambda}_{g}(t) \parallel_{L^{\infty}}\\
& + \parallel \mu_{f}'(t) - \mu_{g}'(t) \parallel_{L^{\infty}}
\leq \xi(t)d
\end{aligned}
\end{equation}
for $t \in [0, \sigma(d)]$. The analogous assertion holds for $t
\leq 0$.
\end{theorem}
\textbf{Proof :} To establish the estimates for the difference to
two solutions at time $t > 0$ we first determine a time interval
on which we get a uniform bound on the $\supp f(t)$ for all
solutions which the initial data are in $D$ and close enough to
$\overset{\circ}{g}$. Define
\begin{align*}
T_{0}(f) : &= \sup \{ t \in ]0, \min(T_{f}, T_{g})[ \, | \quad
\text{such that}\quad 0 \leq s \leq t,\\
& \qquad \parallel e_{f}(s) - e_{g}(s) \parallel_{L^{\infty}} +
\parallel e^{2\lambda_{f}}e_{f}^{2} - e^{2\lambda_{g}}e_{g}^{2}
\parallel_{L^{\infty}}(s)\\
& \qquad \parallel \dot{\lambda}_{f}(s) - \dot{\lambda}_{g}(s)
\parallel_{L^{\infty}} + \parallel \mu'_{f}(s) - \mu'_{g}(s)
\parallel_{L^{\infty}} \leq 1 \}.
\end{align*}
Notice that for $d$ small enough, say $d < \varepsilon_{1}$ for a
suitable defined $\varepsilon_{1} > 0$, the estimate defining
$T_{0}(f)$ holds at $t = 0$ so that by continuity, $T_{0}(f) > 0$.
For a solution $(f, \lambda_{f}, \mu_{f}, e_{f})$ with $d <
\varepsilon_{1}$ the following estimates for the characteristics
hold on $[0, T_{0}(f)[$:
\begin{align*}
& \mid \dot{x}(s) \mid \leq 1,\\
\mid \dot{v}(s) \mid &\leq C(\parallel \dot{\lambda}_{f}(s)
\parallel_{L^{\infty}} + \parallel \mu'_{f}(s)
\parallel_{L^{\infty}} + \parallel e_{f}(s)
\parallel_{L^{\infty}})(1 + \mid v(s) \mid)\\
&\leq C(1 + \parallel \dot{\lambda}_{f}(s)
\parallel_{L^{\infty}} + \parallel \mu'_{f}(s)
\parallel_{L^{\infty}} + \parallel e_{f}(s)
\parallel_{L^{\infty}})(1 + \mid v(s) \mid)
\end{align*}
with $C = C(q)$ being a constant. Via the Gronwall inequality this
implies that
\begin{equation} \label{eq:3.8}
\supp f(t) \subset \{ (\tilde{x}, v) \in \mathbb{R}^{6} \, | \quad
\mid \tilde{x} \mid \leq r_{0} + t, \quad \mid v \mid \leq
U_{g}(t) \}
\end{equation}
for $t \in [0, T_{0}(f)[$, where
\begin{equation*}
U_{g} := (1 + u_{0})\exp\left( C\int_{0}^{t}(1 + \parallel
\dot{\lambda}_{g}(s)
\parallel_{L^{\infty}} + \parallel \mu'_{g}(s)
\parallel_{L^{\infty}} + \parallel e_{g}(s)
\parallel_{L^{\infty}})ds \right).
\end{equation*}
Let $C$ be a continuous, increasing function on $[0, T_{g}[$ which
depends only on $(g, \lambda_{g}, \mu_{g}, e_{g})$. Then we obtain
the following estimates on $[0, T_{0}(f)[$:
\begin{equation} \label{eq:3.9}
\begin{aligned}
&\parallel \rho_{f}(t) - \rho_{g}(t) \parallel_{L^{\infty}} +
\parallel p_{f}(t) - p_{g}(t) \parallel_{L^{\infty}} +
\parallel k_{f}(t) - k_{g}(t) \parallel_{L^{\infty}}\\
&+ \parallel M_{f}(t) - M_{g}(t) \parallel_{L^{\infty}} \leq C(t)
(\parallel f(t) - g(t) \parallel_{L^{\infty}} + \parallel
e^{2\lambda_{f}(t)}e_{f}^{2}(t) - e^{2\lambda_{g}(t)}e_{g}^{2}(t)
\parallel_{L^{\infty}})
\end{aligned}
\end{equation}
where the matter quantity
\begin{equation} \label{eq:3.10}
k(t, \tilde{x}) := \int_{\mathbb{R^{3}}}\frac{\tilde{x} .
v}{r}f(t, \tilde{x}, v)dv
\end{equation}
comes from the following part of the Einstein equations
\cite{noundjeu2}:
\begin{equation} \label{eq:3.11}
\dot{\lambda}(t, \tilde{x}) = - 4\pi r(e^{\lambda + \mu}k)(t,
\tilde{x})
\end{equation}
Now, we have:
\begin{align*}
&\left| \frac{m_{f}(t, r)}{r} - \frac{m_{g}(t, r)}{r} \right|, \,
\left| \frac{m_{f}(t, r)}{r^{2}} - \frac{m_{g}(t, r)}{r^{2}}
\right|\\
&\leq C(t)(\parallel f(t) - g(t) \parallel_{L^{\infty}} +
\parallel e^{2\lambda_{f}(t)}e_{f}^{2}(t) - e^{2\lambda_{g}(t)} e_{g}^{2}(t)
\parallel_{L^{\infty}}).
\end{align*}
The latter is obtained using the following equality deduced from
the integration of (\ref{eq:2.5})on $[r_{0} + t, r]$:
\begin{equation*}
e^{\lambda(t, r)}e(t, r) = \left( \frac{r_{0} + t}{r}
\right)^{2}e^{\lambda(t, r_{0} + t)}e(t, r_{0} + t), \quad r \in
[r_{0} + t, + \infty[.
\end{equation*}
Next as the second step, we derive an estimate for the time
evolution of $f - g$ along the characteristics $Z_{f} = (X_{f},
V_{f})$ corresponding to $f$ (see \cite{noundjeu2}, Proposition
3.1). Using the fact that $f$ is constant along these
characteristics, the mean value theorem and since $\mu - \lambda
\leq 1$, $\mu + \lambda \leq 1$ both for $f$ and $g$, one obtains
the following estimate:
\begin{align*}
\left| \frac{d}{ds}(f - g)(s, Z_{f}(s, t, z)) \right| &\leq
C(s)(\parallel \lambda_{f}(s) - \lambda_{g}(s)
\parallel_{L^{\infty}} + \parallel \mu_{f}(s) - \mu_{g}(s)
\parallel_{L^{\infty}})\\
& \qquad + C(s)\parallel e_{f}(s) - e_{g}(s)
\parallel_{L^{\infty}}\\
& \qquad + C(s)(\parallel \dot{\lambda}_{f}(s) -
\dot{\lambda}_{g}(s) \parallel_{L^{\infty}} + \parallel
\mu'_{f}(s) - \mu'_{g}(s) \parallel_{L^{\infty}});
\end{align*}
and integration of the above w.r.t $s$ on $[0, t]$ yields:
\begin{equation} \label{eq:3.12}
\begin{aligned}
\parallel f(t) - g(t) \parallel_{L^{\infty}} &\leq
\parallel \overset{\circ}{f} - \overset{\circ}{g}
\parallel_{L^{\infty}} + C(t)\int_{0}^{t}\parallel \lambda_{f}(s) -
\lambda_{g}(s)\parallel_{L^{\infty}}ds\\
& \qquad + C(t)\int_{0}^{t}\parallel \mu_{f}(s) -
\mu_{g}(s) \parallel_{L^{\infty}}ds\\
& \qquad + C(t)\int_{0}^{t}(\parallel e_{f}(s) -
e_{g}(s)\parallel_{L^{\infty}} + \parallel \dot{\lambda}_{f}(s) -
\dot{\lambda}_{g}(s) \parallel_{L^{\infty}})ds\\
& \qquad + \int_{0}^{t}\parallel \mu'_{f}(s) -
\mu'_{g}(s)\parallel_{L^{\infty}}ds.
\end{aligned}
\end{equation}
Using (\ref{eq:3.9}) and the fact that $\parallel f(t)
\parallel_{L^{\infty}} = \parallel \overset{\circ}{f}
\parallel_{L^{\infty}} \leq \parallel \overset{\circ}{g}
\parallel_{L^{\infty}} + \varepsilon_{1}$, we obtain the estimates
\begin{align*}
&\parallel \rho_{f}(t) \parallel_{L^{\infty}} +
\parallel p_{f}(t) \parallel_{L^{\infty}}\\
& +\parallel k_{f}(t) \parallel_{L^{\infty}} +
\parallel N_{f}(t) \parallel_{L^{\infty}} \leq C(t), \quad t <
T_{0}(f) \tag{5.10'}\\
\text{and}\\
&\left| \frac{m_{f}(t, r)}{r^{2}} \right| \leq C(t), \quad t <
T_{f}, \quad r > 0.
\end{align*}
where the quantity
\begin{equation} \label{eq:3.13}
N(t, \tilde{x}) := \int_{\mathbb{R}^{3}}
\frac{\tilde{x}.v}{\sqrt{1 + v^{2}}}f(t, \tilde{x}, v)dv
\end{equation}
comes from a following part of the Maxwell equations
\cite{noundjeu2}:
\begin{equation} \label{eq:3.14}
\frac{\partial}{\partial t}(e^{\lambda}e) = - \frac{q}{r}e^{\mu}N.
\end{equation}
Using once again the mean value theorem, the inequalities above
and (\ref{eq:3.11}) one obtains:
\begin{equation} \label{eq:3.15}
\begin{aligned}
\parallel \dot{\lambda}_{f}(t) - \dot{\lambda}_{g}(t)
\parallel_{L^{\infty}} &\leq C(t)(\parallel f(t) - g(t)
\parallel_{L^{\infty}} + \parallel e^{2\lambda_{f}(t)}e_{f}^{2}(t)
- e^{2\lambda_{g}(t)}e_{g}^{2}(t) \parallel_{L^{\infty}})\\
& \qquad + C(t)(\parallel \lambda_{f}(t) - \lambda_{g}(t)
\parallel_{L^{\infty}} + \parallel \mu_{f}(t) - \mu_{g}(t)
\parallel_{L^{\infty}}).
\end{aligned}
\end{equation}
Now, (\ref{eq:2.3}) yields $(-re^{-2\lambda})' = 8\pi r^{2}\rho$
and the integration of this on $[0, r]$ gives:
\begin{equation} \label{eq:3.16}
e^{-2\lambda(t, r)} = 1 - \frac{2m(t, r)}{r}.
\end{equation}
So, combining (\ref{eq:2.3}) and (\ref{eq:3.16}), one obtains:
\begin{equation} \label{eq:3.17}
\mu'(t, r) = e^{2\lambda(t, r)}(\frac{m(t, r)}{r^{2}} + 4\pi rp(t,
r))
\end{equation}
and using (\ref{eq:3.17}), (\ref{eq:3.9}) and the fact that
\begin{equation*}
\lim_{r \to + \infty} \mu_{f}(t, r) = \lim_{r \to + \infty}
\mu_{g}(t, r) = 0,
\end{equation*}
we can deduce by integration on $[0, + \infty[$ w.r.t $r$ the
following inequality:
\begin{equation} \label{eq:3.18}
\begin{aligned}
\parallel \mu_{f}(t) - \mu_{g}(t) \parallel_{L^{\infty}} &\leq C(t)
(\parallel f(t) - g(t) \parallel_{L^{\infty}} + \parallel
e^{2\lambda_{f}(t)} - e^{2\lambda_{g}(t)}\parallel_{L^{\infty}})\\
& \qquad + C(t)\parallel e^{2\lambda_{f}(t)}e_{f}^{2}(t) -
e^{2\lambda_{g}(t)}e_{g}^{2}(t) \parallel_{L^{\infty}}.
\end{aligned}
\end{equation}
We have, since $\lambda + \mu \leq 0$ for both $f$ and $g$:
\begin{equation*}
\mid \partial_{s} e^{2\lambda_{f}} - \partial_{s} e^{2\lambda_{g}}
\mid(s, r) \leq C(s)(\parallel e^{2\lambda_{f}(s)} -
e^{2\lambda_{g}(s)} \parallel_{L^{\infty}} + \parallel
\dot{\lambda}_{f}(s) - \dot{\lambda}_{g}(s)
\parallel_{L^{\infty}}).
\end{equation*}
The insertion of (\ref{eq:3.15}) using (\ref{eq:3.18}) in the
inequality above gives, by integration:
\begin{equation} \label{eq:3.19}
\begin{aligned}
\parallel e^{2\lambda_{f}(t)} - e^{2\lambda_{g}(t)}
\parallel_{L^{\infty}} &\leq \parallel e^{2\overset{\circ}{\lambda}_{f}}
- e^{2\overset{\circ}{\lambda}_{g}} \parallel_{L^{\infty}} +
C(t)\int_{0}^{t}\parallel f(s) - g(s) \parallel_{L^{\infty}}ds\\
& \qquad + C(t)\int_{0}^{t}\parallel \lambda_{f}(s) -
\lambda_{g}(s) \parallel_{L^{\infty}}ds\\
& \qquad + C(t)\int_{0}^{t}\parallel \mu_{f}(s) -
\mu_{g}(s) \parallel_{L^{\infty}})ds\\
& \qquad + C(t)\int_{0}^{t}\parallel e^{2\lambda_{f}(s)} -
e^{2\lambda_{g}(s)} \parallel_{L^{\infty}}ds\\
& \qquad + C(t)\int_{0}^{t}\parallel
e^{2\lambda_{f}(s)}e_{f}^{2}(s) - e^{2\lambda_{g}(s)}e_{g}^{2}(s)
\parallel_{L^{\infty}}ds.
\end{aligned}
\end{equation}
Now, to obtain an estimate for the last term on the right hand
side of (\ref{eq:3.19}), we use (\ref{eq:3.14}) and the fact that
since $\mu \leq 0$ and $\lambda + \mu \leq 0$, one gets:
\begin{align*}
&\mid N_{g}(s, r)\mid \leq (r_{0} + t)\parallel \overset{\circ}{g}
\parallel_{L^{\infty}} \leq C(s)\\
&\mid e^{\mu_{f}} - e^{\mu_{g}} \mid(s, r) \leq \mid \mu_{f} -
\mu_{g} \mid(s, r) \quad \text{(Mean value theorem)},
\end{align*}
to deduce by integration that:
\begin{equation} \label{eq:3.20}
\begin{aligned}
\parallel e^{2\lambda_{f}(t)}e_{f}^{2}(t) - e^{2\lambda_{g}(t)}
e_{g}^{2}(t) \parallel_{L^{\infty}} &\leq \parallel
e^{2\overset{\circ}{\lambda}_{f}}\overset{\circ}{e}_{f}^{2} -
e^{2\overset{\circ}{\lambda}_{g}}\overset{\circ}{e}_{g}^{2}
\parallel_{L^{\infty}}\\
& \qquad + \int_{0}^{t}C(s)\parallel f(s) - g(s)
\parallel_{L^{\infty}}ds\\
& \qquad + \int_{0}^{t}C(s)\parallel \mu_{f}(s) - \mu_{g}(s)
\parallel_{L^{\infty}}ds\\
& \qquad + \int_{0}^{t}C(s)\parallel e^{\lambda_{f}(s)}e_{f}(s) -
e^{\lambda_{g}(s)} e_{g}(s) \parallel_{L^{\infty}}.
\end{aligned}
\end{equation}
To deduce an estimate for the last term in the right hand side of
(\ref{eq:3.20}), we use once again (\ref{eq:3.14}) and obtain by
integration:
\begin{equation} \label{eq:3.21}
\begin{aligned}
\parallel e^{\lambda_{f}(t)}e_{f}(t) - e^{\lambda_{g}(t)}e_{g}(t)
\parallel_{L^{\infty}} &\leq \parallel e^{\overset{\circ}{\lambda}_{f}}
\overset{\circ}{e}_{f} - e^{\overset{\circ}{\lambda}_{g}}
\overset{\circ}{e}_{g} \parallel_{L^{\infty}}\\
& \qquad + \int_{0}^{t}C(s)\parallel f(s) - g(s)
\parallel_{L^{\infty}}ds\\
& \qquad + \int_{0}^{t}C(s)\parallel \mu_{f}(s) - \mu_{g}(s)
\parallel_{L^{\infty}}ds.
\end{aligned}
\end{equation}
So, using (\ref{eq:3.21}), (\ref{eq:3.20}), (\ref{eq:3.19}),
(\ref{eq:3.12}) and the Gronwall inequality, one obtains, since $C
\in C([0, T_{g}[)$ depends only on $g$ and can be taken strictly
increasing with $\underset{t \rightarrow T_{g}}{\lim} C(t) = +
\infty$:
\begin{equation} \label{eq:3.22}
\begin{aligned}
&\parallel f(t) - g(t)\parallel_{L^{\infty}} + \parallel
\lambda_{f}(t) - \lambda_{g}(t) \parallel_{L^{\infty}} + \parallel
e^{2\lambda_{f}(t)} - e^{2\lambda_{g}(t)} \parallel_{L^{\infty}}\\
& \qquad + \parallel e^{2\lambda_{f}(t)}e_{f}^{2}(t) -
e^{2\lambda_{g}(t)}e_{g}^{2}(t) \parallel_{L^{\infty}} +
\parallel e^{\lambda_{f}(t)}e_{f}(t) - e^{\lambda_{g}(t)}e_{g}(t)
\parallel_{L^{\infty}} \leq \xi(t)d
\end{aligned}
\end{equation}
and since (\ref{eq:3.17}) implies the estimate:
\begin{align*}
\mid \mu_{f}' - \mu_{g}' \mid &\leq \left( \left| \frac{m_{f}(t, r
)}{r^{2}} \right| + \frac{r}{2}\mid p_{f}(t, r) \mid \right)\mid
e^{2\lambda_{f}} - e^{2\lambda_{g}} \mid(t,r)\\
&\qquad + e^{2\lambda_{g}(t, r)}\left| \frac{m_{f}(t, r )}{r^{2}}
- \frac{m_{g}(t, r )}{r^{2}} \right| + \frac{r}{2}\mid p_{f} -
p_{g} \mid(t, r),
\end{align*}
one deduce from (\ref{eq:3.22}) that:
\begin{equation*}
\parallel e_{f}(t) - e_{g}(t) \parallel_{L^{\infty}} + \parallel
\dot{\lambda}_{f}(t) - \dot{\lambda}_{g}(t) \parallel_{L^{\infty}}
+ \parallel \mu'_{f}(t) - \mu'_{g}(t) \parallel_{L^{\infty}} \leq
\xi(t)d
\end{equation*}
where $\xi \in C([0, T_{g}[)$ depends only on $g$, strictly
increasing, $\xi(0) > 0$, and $\underset{t \rightarrow
T_{g}}{\lim} \xi(t) = + \infty$. Define
\begin{equation*}
\varepsilon := \min \left\{ \varepsilon_{1}, \frac{1}{2\xi(0)}
\right\}; \quad \sigma(\beta) := \xi^{-1}\left( \frac{1}{2\beta}
\right); \quad \beta \in ]0, \varepsilon[.
\end{equation*}
Then $\sigma \in C(]0, \varepsilon[)$ is positive, strictly
decreasing, and $\underset{\beta \rightarrow 0}{\lim}
\sigma(\beta) = T_{g}$. Let $(\overset{\circ}{f},
\overset{\circ}{\lambda}, \overset{\circ}{e}) \in D$ be such that
$0 < d < \varepsilon$. Then on the interval $[0, \min\{ \sigma(d),
T_{0}(f)\}]$ the following estimate holds since $\xi$ is strictly
increasing:
\begin{align*}
A(t) := &\parallel e_{f}(t) - e_{g}(t) \parallel_{L^{\infty}} +
\parallel e^{2\lambda_{f}}e_{f}^{2}(t) - e^{2\lambda_{g}}e_{g}^{2}(t)
e_{g}^{2}(t) \parallel_{L^{\infty}}\\
+ &\parallel \dot{\lambda}_{f}(t) - \dot{\lambda}_{g}(t)
\parallel_{L^{\infty}} + \parallel \mu'_{f}(t) - \mu'_{g}(t)
\parallel_{L^{\infty}} \leq \xi(t)d < \xi(\sigma(d))d =
\frac{1}{2d}d = \frac{1}{2}.
\end{align*}
Thus,
\begin{equation} \label{eq:3.23}
A(t) < \frac{1}{2}, \quad \text{for} \quad t < \sigma(d).
\end{equation}
Assume $T_{f} \leq \sigma(d)$. Then by definition of $T_{0}(f)$
and (\ref{eq:3.23}) we obtain the identity $T_{0}(f) = \min
\{T_{f}, T_{g} \} = T_{f}$, in particular the estimate
\begin{equation*}
\mid v \mid \leq U_{g}(\sigma(d)) < \infty,
\end{equation*}
holds for all $(\tilde{x}, v) \in \supp f(t)$ and $t \in [0,
T_{f}[$. Since $T_{f} \leq \sigma(d) < \infty$, this is a
contradiction to theorem 3.1, and we have shown that $T_{f} >
\sigma(d)$. Furthermore, (\ref{eq:3.23}), implies that $T_{0}(f)
> \sigma(d)$ so that the estimates which were established on the
interval $[0, T_{0}(f)[$ hold on $[0, \sigma(d)]$, and the proof
is complete.

Next, we use the above theorem and the following formula
\begin{equation*}
\lambda'(t, r) = e^{2\lambda(t, r)}\left( -\frac{m(t, r)}{r^{2}} +
\frac{r}{2}\rho(t, r) \right),
\end{equation*}
to obtain as it is proved in detail in \cite{noundjeu3}, the
\begin{corollary} \label{C:3.1}
Let $(\overset{\circ}{f}, \overset{\circ}{\lambda},
\overset{\circ}{e}) \in D$ be such that
\begin{equation*}
d = \parallel \overset{\circ}{f} - \overset{\circ}{g}
\parallel_{L^{\infty}} < \varepsilon.
\end{equation*}
Then the following inequalities hold on $[0, \sigma(d)]$:
\begin{equation*}
\quad \supp f(t) \subset \{ (\tilde{x}, v) \in \mathbb{R}^{6} \, |
\, \mid \tilde{x} \mid \leq r_{0} + t, \quad \mid v \mid \leq
U_{g}(t) \}
\end{equation*}
\begin{equation} \label{eq:3.24}
\parallel \lambda'_{f}(t) - \lambda'_{g}(t)
\parallel_{L^{\infty}} + \parallel \dot{\mu}_{f}(t) - \dot{\mu}_{g}(t)
\parallel_{L^{\infty}} \leq S(t)d
\end{equation}
\begin{equation} \label{eq:3.25}
\quad \frac{1}{r}(\mid \dot{\lambda}_{f}(t) - \dot{\lambda}_{g}(t)
\mid + \mid \lambda'_{f}(t) - \lambda'_{g}(t) \mid + \mid
\mu'_{f}(t) - \mu'_{g}(t) \mid)(t, r) \leq S(t)d
\end{equation}
\begin{equation} \label{eq:3.26}
\mid Z_{f} - Z_{g} \mid(t, 0, z) \leq S(t)d, \quad \text{for}
\quad z \in \supp \overset{\circ}{f} \cup \supp \overset{\circ}{g}
\end{equation}
\begin{equation} \label{eq:3.27}
\parallel \Gamma_{\beta \gamma f}^{\alpha}(t) -
\Gamma_{\beta \gamma g}^{\alpha}(t)
\parallel_{L^{\infty}} \leq S(t)d
\end{equation}
\begin{equation} \label{eq:3.28}
\parallel R_{\alpha}\, ^{\delta}_{, \beta \gamma
f}(t) - R_{\alpha}\, ^{\delta}_{, \beta \gamma g}(t)
\parallel_{L^{\infty}} \leq S(t)d,
\end{equation}
where $\sigma$ and $\varepsilon$ are introduced as in Theorem 3.2
and $S \in C([0, T_{g}[)$ is a positive increasing function.
\end{corollary}
Using once again Theorem 3.2 and Lemma 3.1 in \cite{noundjeu2},
one deduces the
\begin{corollary} \label{C:3.2}
With the assumption in Theorem 3.2, let $\overset{\circ}{g} \in
C^{2}(\mathbb{R}^{6})$. Then the following
\begin{equation*}
\parallel \partial_{z} f(t) - \partial_{z} g(t) \parallel_{L^{\infty}}
\leq S(t)(\parallel \overset{\circ}{f} - \overset{\circ}{g}
\parallel_{L^{\infty}} + \parallel \partial_{z} \overset{\circ}{f}
- \partial_{z} \overset{\circ}{g} \parallel_{L^{\infty}}), \quad t
\in [0, \sigma(d)]
\end{equation*}
holds for initial  data $(\overset{\circ}{\lambda},
\overset{\circ}{e}, \overset{\circ}{f}) \in D$, with $d \leq
\varepsilon$, and $\parallel \partial_{z} \overset{\circ}{f} -
\partial_{z} \overset{\circ}{g} \parallel_{L^{\infty}} < 1$, where
$\varepsilon$, $\sigma$ and $d$ are as in Theorem 3.2 and the
nonnegative, increasing function $S \in C([0, T_{g}[)$ being as in
the Corollary 3.1.
\end{corollary}
\section{Global existence for small initial data}
This section is concerned with the proof of a global existence
theorem of solutions when the initial data and the charge of
particle are sufficiently small. We also observe that in the
corresponding spacetime, all geodesics are complete and then the
obtained spacetime is complete. Here is the main result of this
section:
\begin{theorem} \label{T:4.1}
For all $r_{0} > 0$, $u_{0} > 0$ and $\Lambda > 0$ there exists
$\varepsilon > 0$ such that if $(\lambda, \mu, e, f)$ is the
maximal solution of the asymptotically flat, spherically symmetric
EVM system with data $(\overset{\circ}{\lambda},
\overset{\circ}{e}, \overset{\circ}{f})$, satisfying
\begin{equation*}
\supp \overset{\circ}{f} \subset B(r_{0}) \times B(u_{0}), \quad
\parallel \overset{\circ}{f}
\parallel_{L^{\infty}} < \varepsilon, \parallel \overset{\circ}{\lambda}
\parallel_{L^{\infty}} \leq \Lambda,
\end{equation*}
$(\overset{\circ}{\lambda}, \overset{\circ}{e})$ being a regular
solution of the constraints, then the solution exists globally in
$t$. Moreover, it satisfies condition (FS) stated below on
$\mathbb{R}$ with $\delta = 1$ and some constant $\gamma > 0$,
\begin{align*}
\parallel \rho(t) \parallel_{L^{\infty}},
\parallel p(t) \parallel_{L^{\infty}},
\parallel k(t) \parallel_{L^{\infty}},
\parallel M(t) \parallel_{L^{\infty}} &= \mathcal{O}((1 + \mid t
\mid)^{-3})\\
\parallel \lambda(t) \parallel_{L^{\infty}},
\parallel \mu(t) \parallel_{L^{\infty}}, &= \mathcal{O}((1 + \mid t \mid)^{-1})\\
\parallel \Gamma_{\beta \gamma}^{\alpha}(t)
\parallel_{L^{\infty}},
\parallel N(t) \parallel_{L^{\infty}},
\parallel e(t) \parallel_{L^{\infty}} &= \mathcal{O}((1 + \mid t
\mid)^{-2})\\
\parallel R_{\alpha, \delta \gamma}^{\,\beta}(t)
\parallel_{L^{\infty}} &= \mathcal{O}((1 + \mid t \mid)^{-3})
\end{align*}
for $t \in \mathbb{R}$, and the geodesics are defined on
$\mathbb{R}$.
\end{theorem}
Before giving a sketch of the proof of this result (reader can
refer to \cite{noundjeu3} to get more details), we need the
following result obtained by distinguishing the cases $r \leq
r_{0}$ and $r \geq r_{0}$:
\begin{proposition} \label{P:4.1}
Let $(\overset{\circ}{f}, \overset{\circ}{\lambda},
\overset{\circ}{e}) \in D$. Then the following inequalities hold:
\begin{align*}
\mid \overset{\circ}{e}(r) \mid &\leq C\parallel
\overset{\circ}{f} \parallel_{L^{\infty}}\\
\mid \overset{\circ}{\mu}(r) \mid, \, \mid m(0, r) \mid &\leq
C\parallel \overset{\circ}{f}
\parallel_{L^{\infty}}(1 + \parallel \overset{\circ}{f}
\parallel_{L^{\infty}})
\end{align*}
for all $r \geq 0$, where $C$ is a constant with depends on
$r_{0}$ and $\Lambda$. and $m(0, r)$ is deduced from
(\ref{eq:3.4}) setting $r = 0$.
\end{proposition}
\subsection{The free-streaming (FS) condition}
Take a regular solution $(f, \lambda, \mu, e)$ which exists on a
time interval $[0, T[$. For $\delta \in ]0, 1]$ and $\gamma > 0$
we consider the following decay condition on an interval $[0, T'[
\subset [0, T[$:
\begin{equation*}
(FS) \quad
\begin{cases}
\mid \dot{\lambda}(t, r) \mid + \mid \dot{\mu}(t, r) \mid + \mid
\lambda'(t, r) \mid + \mid \mu'(t, r) \mid \leq \gamma (1 + t
)^{-1 - \delta}\\
\mid M(t, r) \mid + \left| \frac{e(t, r)}{r} \right| + \mid H(t,
r) \mid
\leq \gamma (1 + t)^{-2 - \delta}\\
\frac{1}{r}(\mid \dot{\lambda}(t, r) \mid + \mid \mu'(t, r) \mid +
\mid \lambda'(t, r) \mid) \leq \gamma (1 + t)^{-2 - \delta}\\
\mid e(t, r) \mid \leq \gamma (1 + t)^{-1 - \delta}
\end{cases}
\end{equation*}
for $t \in [0, T'[$ and $r \geq 0$; where
\begin{equation*}
H = e^{-2\lambda}\left( \mu'' + (\mu' - \lambda')(\mu' +
\frac{1}{r}) \right) - e^{-2\mu}(\ddot{\lambda} +
\dot{\lambda}(\dot{\lambda} - \dot{\mu})).
\end{equation*}
The following result, obtained by using the Gronwall inequality,
shows that under such an assumption, the momenta cannot grow very
much:
\begin{lemma} \label{L:4.1}
Let $\delta \in ]0, 1]$ and $u_{0} > 0$. If $\gamma$ is
sufficiently small and if $(f, \lambda, \mu, e)$ is a solution
which satisfies (FS) on an interval $[0, T'[$, then every solution
of the characteristic system satisfies the estimate
\begin{equation*}
\mid V(t, 0, \tilde{x}, v) \mid \leq u_{0} + 1, \, (\tilde{x}, v)
\in \mathbb{R}^{3} \times B(u_{0}), \, t \in [0, T'[.
\end{equation*}
\end{lemma}
Next, using Lemma 3.1 in \cite{noundjeu2}, one obtains as in the
uncharged case, that any solution which satisfies (FS) allows the
matter quantity $\rho$ to behave like $t^{-3}$. To obtain the
latter, it suffices to show that $\det (\partial_{v}X(0, t,
\tilde{x}, v))^{-1}$ decays like $t^{-1}$ and one deduces:
\begin{equation*}
\int_{\mathbb{R}^{3}}\sqrt{1 + v^{2}}f(t, \tilde{x}, v)dv \leq
C_{1}t^{-3}, \quad t \in ]0, T'[.
\end{equation*}
Since
\begin{equation*}
\frac{1}{2}e^{2\lambda(t, r)}e^{2}(t, r) \leq
\frac{e^{2}}{2}t^{-3},
\end{equation*}
we obtain the desired result that is:
\begin{lemma} \label{L:4.2}
Let $\delta \in ]0, 1]$ and $r_{0}, u_{0}, C_{0} > 0$. Then there
exists constants $\gamma > 0$ and $C_{1} > 0$ such that if
solution $(f, \lambda, \mu, e)$ satisfies the free-streaming
condition (FS) on an interval $[0, T'[$, $\supp f(0) \subset
B(r_{0}) \times B(u_{0})$ and
\begin{equation*}
\parallel \overset{\circ}{f} \parallel_{L^{\infty}} +
\parallel \overset{\circ}{\lambda} \parallel_{L^{\infty}}
+ \parallel \overset{\circ}{\mu} \parallel_{L^{\infty}} +
\parallel \overset{\circ}{e} \parallel_{L^{\infty}} \leq C_{0},
\end{equation*}
then
\begin{equation*}
\parallel \rho(t) \parallel_{L^{\infty}} \leq C_{1}t^{-3}, \quad t
\in ]0, T'[.
\end{equation*}
\end{lemma}
\begin{remark} \label{R:4.1}
We are not surprised by the fact that $\rho$ decays like $t^{-3}$,
since even in the Minkowski spacetime, free particles starting in
a compact set spread out linearly with time and the associated
density decays like $t^{-3}$ as $t \rightarrow \infty$ (see
\cite{rendall1}).
\end{remark}
\subsection{Decay of the fields}
Let $(f, \lambda, \mu, e)$ be a solution so that (FS) is
satisfied. From the following estimates
\begin{align*}
\mid e(t, r) \mid &\leq C\frac{1}{r^{2}}\int_{0}^{r}s^{2}\mid M(t,
s)\mid ds \leq C(1 + t)^{-2}\\
\mid e(t, r) \mid &\leq C\frac{1}{r^{3}}\int_{0}^{r}s^{2}\mid M(t,
s) \mid ds \leq C(1 + t)^{-2}\\
\mid H(t, r) \mid &\leq  C(1 + t)^{-2}
\end{align*}
one deduces that the decay of $\rho$ implies the decay assumptions
stated in (FS):
\begin{lemma} \label{L:4.3}
Let $r_{0}, C_{0}, C_{1} > 1$. Then there exists a constant
$\gamma > 0$ such that if $(f, \lambda, \mu, e)$ is a solution on
$[0, T[$, satisfying the estimates
\begin{equation*}
\parallel \rho(t) \parallel_{L^{\infty}} \leq C_{1}(1 + t)^{-3},
\quad t \in [0, T'[,
\end{equation*}
and
\begin{equation*}
\sup \{ \mid \tilde{x} \mid | \, (\tilde{x}, v) \in \supp f(0) \}
\leq r_{0}, \quad \parallel \rho(t) \parallel_{L^{\infty}} \leq
C_{0}
\end{equation*}
for some $T' \in ]0, T[$, then $(f, \lambda, \mu, e)$ satisfies
the free-streaming condition (FS) on the interval $[0, T'[$ with
the parameters $\delta = 1$ and $\gamma$.
\end{lemma}
\subsection{Proof of Theorem 4.1}
Let $r_{0}$, $u_{0} > 0$ and $\Lambda > 0$ be fixed. Take
$(\overset{\circ}{f}, \overset{\circ}{\lambda},
\overset{\circ}{e}) \in D$, where $D$ is the set of initial data
introduced in Sect.3. Using Proposition 4.1, we observe that if
$\overset{\circ}{f}$ is small in the $L^{\infty}$-norm, then so
are $\overset{\circ}{\mu}$ and $\overset{\circ}{e}$. We choose
$\varepsilon > 0$ small enough in such a way that for all
nonnegative, spherically symmetric initial data
$\overset{\circ}{f} \in C_{c}^{\infty}(\mathbb{R}^{6})$ with
\\ $\supp \overset{\circ}{f} \subset B(r_{0}) \times B(u_{0})$, and
$\parallel \overset{\circ}{f} \parallel_{L^{\infty}} <
\varepsilon$ the estimates
\begin{align*}
&8\pi \int_{\mid y \mid \leq r}\int_{\mathbb{R}^{3}}\sqrt{1 +
v^{2}}\overset{\circ}{f}(y, v)dvdy < r, \quad r \geq 0,\\
&\parallel \overset{\circ}{\lambda} \parallel_{L^{\infty}} \leq
\Lambda, \quad \parallel \overset{\circ}{f} \parallel_{L^{\infty}}
+ \parallel \overset{\circ}{\mu} \parallel_{L^{\infty}} +
\parallel \overset{\circ}{e} \parallel_{L^{\infty}} \leq 1
\end{align*}
hold. Using Theorem 3.1, we have for such initial data a local
solution on some right maximal existence interval $[0, T[$ and we
can choose $C_{0} = 1$ when applying Lemma 4.2 and Lemma 4.3. Take
$g = \lambda_{g} = \mu_{g} = e_{g} = 0$, and $T_{g} = 1$. Applying
Theorem 3.2, there exists $\varepsilon > 0$, a positive decreasing
function $\xi \in C([0, 1])$ and a positive decreasing function
$\sigma \in C(]0, \varepsilon])$ such that $\underset{\beta
\rightarrow 0}{\lim} \sigma(\beta) = 1$, and for any solution $(f,
\lambda_{f}, \mu_{f}, e_{f})$ with $d = \parallel
\overset{\circ}{f} \parallel_{L^{\infty}} < \varepsilon$, and the
estimates below
\begin{align*}
&\parallel f(t) \parallel_{L^{\infty}}, \, \parallel
e^{2\lambda_{f}(t)}
\parallel_{L^{\infty}}, \, \parallel e_{f}(t) \parallel_{L^{\infty}} \leq
\xi(t)\varepsilon, \quad t \in [0, \sigma(\varepsilon)]\\
\text{and then}\\
&\parallel \rho(t) \parallel_{L^{\infty}} \leq C\xi(t)\varepsilon,
\quad t \in [0, \sigma(\varepsilon)],
\end{align*}
where $C = C(u_{0})$ is a constant. So, we can choose
$\varepsilon$ small enough to have $CL\varepsilon \leq 1$, where
$L := \underset{t \in [0, 1]}{\sup} \xi(t)$ and obtain a solution
$(f, \lambda, \mu, e)$ which is defined on the interval $[0, 1]$,
with
\begin{equation*}
\parallel \rho(t) \parallel_{L^{\infty}} \leq 1, \quad t \in [0,
1].
\end{equation*}
Take $\delta = \frac{1}{2}$ and choose a corresponding $\gamma >
0$ such that Lemma 4.2 and \\ Lemma 4.3 hold. Let $C_{\gamma}$ be
the constant corresponding to $\gamma$ and define
\begin{equation*}
C^{\ast} := 8(C_{\gamma} + 1)
\end{equation*}
and we take for instance $T_{1} =
\frac{4\gamma_{C^{\ast}}^{2}}{\gamma^{2}} + 1$ to have
\begin{equation*}
\gamma_{C^{\ast}}(1 + t)^{-1} \leq \frac{\gamma}{2}(1 +
t)^{\frac{-1}{2}}, \quad \text{for} \, t \geq T_{1}.
\end{equation*}
Using Theorem 3.2 and Corollary 3.1 with $g = 0$, we can choose
$\varepsilon$ such that the solution $(f, \lambda, \mu, e)$ exists
on $[0, T_{1}]$ and on this interval the condition (FS) with
parameters $\delta = \frac{1}{2}$ and $\gamma$ considered above,
provided $\parallel \overset{\circ}{f} \parallel_{L^{\infty}} <
\varepsilon$. Set
\begin{equation*}
T_{2} := \sup \{ t \in [0, T[ | \, (f, \lambda, \mu, e) \,
\text{satisfies} \, (FS) \, \text{on} \, [0, t] \}.
\end{equation*}
Then by definition $T_{2} > T_{1}$, and using Lemma 4.2
\begin{equation*}
\parallel \rho(t) \parallel_{L^{\infty}} \leq C_{\gamma}t^{-3},
\quad t \in ]0, T_{2}[,
\end{equation*}
and we use the fact that $\parallel \rho(t)
\parallel_{L^{\infty}} \leq 1$ for $t \in [0, 1]$, to establish
the following inequality:
\begin{equation*}
\parallel \rho(t) \parallel_{L^{\infty}} \leq C^{\ast}(1 +
t)^{-3}, \quad t \in [0, T_{2}[.
\end{equation*}
Note that the inequality above is obtained, distinguishing the
cases $0 \leq t \leq 1$ and $1 < t < T_{2}$. Now, using Lemma 4.3,
the free-streaming condition (FS) holds with the parameters
$\delta = 1$ and $\gamma_{C^{\ast}}$, and with the choice of
$T_{1}$, (FS) holds again on $[T_{1}, T_{2}[$ with parameters
$\frac{\gamma}{2}$ and $\delta = \frac{1}{2}$. By the construction
of $T_{2}$ we obtain $T_{2} = T$. We deduce from Lemma 4.1
\begin{equation*}
\supp f(t) \subset \mathbb{R} \times B(u_{0} + 1), \quad t \in [0,
T[
\end{equation*}
and using Theorem 3.1, we conclude that $T = \infty$.

Note that the decay estimates of $p(t)$, $k(t)$, $M(t)$ and $N(t)$
come with the proof. So, we just have to estimate the metric, the
Christoffel symbols and the Riemann curvature tensor. From
\begin{equation*}
\lambda(t, r) = -\int_{r}^{+\infty}\lambda'(t, s)ds = -
\int_{r}^{+\infty}e^{2\lambda(t, s)}\left( -\frac{m(t, s)}{s^{2}}
+ 4\pi s\rho(t, s) \right)ds
\end{equation*}
and
\begin{align*}
\mid \lambda(t, r) \mid &= \left| \lambda(0, r) + \int_{0}^{t}
\dot{\lambda}(s, r)ds \right|\\
&\leq \parallel \lambda(0) \parallel_{L^{\infty}} +
C\int_{0}^{\infty}(1 + s)^{-2}ds\\
&\leq C
\end{align*}
we deduce the following:
\begin{align*}
\mid \lambda(t, r) \mid &\leq C\int_{0}^{r_{0} + t}\frac{m(t,
s)}{s^{2}}ds + C\int_{r_{0} + t}^{\infty}\frac{M}{s^{2}}ds + C(1 +
t)^{-3}\int_{0}^{r_{0} + t}sds\\
&\leq C(1 + t)^{-1}, \, r_{0} \geq 0, \, t \geq 0,
\end{align*}
where $M > 0$ is the A.D.M mass of the solution. The estimates for
$\mu$ are similar. The estimate of Christoffel symbols is based on
the following:
\begin{equation*}
\frac{1 - e^{-2\lambda(t, r)}}{r} = \frac{2m(t, r)}{r^{2}} \leq
8\pi \int_{0}^{r_{0} + t}\rho(t, s)ds \leq C(1 + t)^{-2},
\end{equation*}
and the estimate of Riemann curvature tensor is deduced from the
above. Now, from the above decay, we deduce that
\begin{equation*}
\lim_{t \to + \infty} \lambda(t, r) = \lim_{t \to + \infty} \mu(t,
r) = 0
\end{equation*}
and since we get
\begin{equation*}
\parallel e(t) \parallel_{L^{\infty}} \leq \sqrt{2}
\parallel \rho(t) \parallel_{L^{\infty}},
\end{equation*}
one concludes that $e(t, r) \underset{t \rightarrow +
\infty}{\rightarrow 0}$. So, the solution $(f, \lambda, \mu, e)$
is asymptotically flat in time coordinate. To end the proof of
Theorem 4.1, we notice that all geodesics are complete. To see the
latter, one proceeds exactly as in the uncharged case (we refer to
\cite{rein1}). Thus the proof of Theorem 4.1 is complete.
\section{The Restricted Regularity Theorem}
In this section, we aim to prove that a solution may be extended
whenever $f$ vanishes in a neighborhood of the center of symmetry.
Take $\overset{\circ}{f} \geq 0$, spherically symmetric,
$C^{\infty}$, compactly supported so that (\ref{eq:3.1}) is
satisfied. Let $(\overset{\circ}{\lambda}, \overset{\circ}{\mu},
\overset{\circ}{e})$ be a solution of the constraints. Then
Theorem 3.1 launches the regular solution $(f, \lambda, \mu, e)$
of the initial boundary value problem on $[0, T] \times
\mathbb{R}^{6}$, for some $T > 0$.
\begin{theorem} \label{T:5.1}
Let $(\overset{\circ}{f}, \overset{\circ}{\lambda},
\overset{\circ}{\mu}, \overset{\circ}{e})$ and $T$ be as above
($T$ finite). Assume there exists $\varepsilon > 0$ such that
\begin{equation*}
f(t, \tilde{x}, v) = 0 \quad \text{if} \quad 0 \leq t < T \quad
\text{and} \quad  \mid \tilde{x} \mid \leq \varepsilon.
\end{equation*}
Then $(f, \lambda, \mu, e)$ extends to a regular solution on $[0,
T'[$ for some $T' > T$.
\end{theorem}
\textbf{Proof:} Define
\begin{equation} \label{eq:5.1}
P(t) := \sup \{ \mid v \mid: \, (\tilde{x}, v) \in \supp f(t) \}
\end{equation}
By Theorem 3.1, it suffices to show that $P(t)$ is bounded on $[0,
T[$. To do so, we introduce the following notation:
\begin{equation*}
u := \mid v \mid, \quad w := r^{-1}\tilde{x} . v, \quad L :=
r^{2}u^{2} - (\tilde{x} . v)^{2}.
\end{equation*}
Note that the square root of $L$ is called the angular momentum.
Since $L$ is constant along the characteristics, we obtain:
\begin{equation} \label{eq:5.2}
\dot{L} = 0
\end{equation}
\begin{equation} \label{eq:5.3}
\dot{r} = e^{\mu - \lambda}\frac{w}{\sqrt{1 + u^{2}}}
\end{equation}
\begin{equation} \label{eq:5.4}
\dot{w} = \frac{L}{r^{3}\sqrt{1 + u^{2}}}e^{\mu - \lambda} -
w\dot{\lambda} - \mu'e^{\mu - \lambda}\sqrt{1 + u^{2}} +
qe^{\lambda + \mu}e.
\end{equation}
Now, the values of any solution $f$ of the Vlasov equation are
conserved along characteristics. So,
\begin{equation*}
0 \leq f \leq \sup \overset{\circ}{f} \leq C
=C(\overset{\circ}{f}, \varepsilon, T).
\end{equation*}
Next, multiplying the Vlasov equation by $\sqrt{1 + v^{2}}$ and
integrating in $v$, using (\ref{eq:3.14}) to show that
\begin{equation*}
1/2\partial_{t}(e^{2\lambda}e^{2}) = - qe^{\lambda + \mu}eN/r,
\quad \partial_{t} := \frac{\partial}{\partial t}
\end{equation*}
and the definition of $\dot{\lambda}$ plus the relation $\lambda'
+ \mu' = 4\pi re^{2\lambda}(\rho + p)$, yields the following
conservation law:
\begin{equation} \label{eq:5.5}
\partial_{t}\rho + \underset{\tilde{x}}{\div} \left( e^{\mu - \lambda}
\int_{\mathbb{R}^{3}} vfdv \right) = 0.
\end{equation}
So, the integration of (\ref{eq:5.5}) in $\tilde{x}$-variable
yields, since $f$ is compactly supported:
\begin{equation*}
\int_{\mathbb{R}^{3}}\rho(t, \tilde{x}) d\tilde{x} =
\int_{\mathbb{R}^{3}}\rho(0, \tilde{x}) d\tilde{x} = C
\end{equation*}
and one deduces:
\begin{equation*}
0 \leq m(t, r) \leq \int_{\mathbb{R}^{3}}\rho(t,
\tilde{x})d\tilde{x} \leq C, \quad r \geq 0.
\end{equation*}
Also, we have $0 \leq L \leq C$ on the support of $f$. Thus
\begin{equation} \label{eq:5.6}
u^{2} = w^{2} + L/r^{2} \leq w^{2} + C/\varepsilon^{2} \leq w^{2}
+ C.
\end{equation}
The inequality (\ref{eq:5.6}) shows that we need to obtain a bound
for $w$. To do this, we introduce
\begin{align*}
P_{i}(t) &:= \inf \{ w: \exists \tilde{x}, v \in \mathbb{R}^{3} \,
\text{with} \, f(t, \tilde{x}, v) \neq 0 \, \text{and} \, w =
r^{-1}\tilde{x} . v \}\\
P_{s}(t) &:= \sup \{ w: \exists \tilde{x}, v \in \mathbb{R}^{3} \,
\text{with} \, f(t, \tilde{x}, v) \neq 0 \, \text{and} \, w =
r^{-1}\tilde{x} . v \}.
\end{align*}
Since
\begin{equation} \label{eq;5.7}
\measure \{ v: (\tilde{x}, v) \in \supp f(t) \} \leq \pi
C\varepsilon^{-2}(P_{s}(t) - P_{i}(t)), \quad \mid \tilde{x} \mid
\geq \varepsilon
\end{equation}
one deduces that if $P_{i}(t)$ and $P_{s}(t)$ are bounded, so is
$P(t)$. To do so, we concentrate on the characteristic equation
for $w$. Using (\ref{eq:3.17}), one obtains:
\begin{equation} \label{eq:5.8}
\begin{aligned}
\dot{w} &= \frac{L}{r^{3}\sqrt{1 + u^{2}}}e^{\mu - \lambda} -
r^{-2}me^{\lambda + \mu}\sqrt{1 + u^{2}} + 4\pi re^{\mu +
\lambda}(wk - p\sqrt{1 + u^{2}})\\
& \qquad + qe^{\mu + \lambda}e.
\end{aligned}
\end{equation}
As it is shown in \cite{rein2}, the first term in the right hand
side of (\ref{eq:5.8}) is bounded on the support of $f$. The same
is true for the second term:
\begin{equation} \label{eq:5.9}
0 \leq r^{-2} me^{\mu + \lambda}\sqrt{1 + u^{2}} \leq
\varepsilon^{-2}C\sqrt{C + w^{2}} \leq C\sqrt{C + w^{2}}.
\end{equation}
We now focus on a bound of the contribution of electric field. The
integration of (\ref{eq:2.5}) in $r$-variable on $[0, r]$ yields:
\begin{equation} \label{eq:5.10}
e(t, r) = \frac{q}{r^{2}}e^{-\lambda(t,
r)}\int_{0}^{r}s^{2}e^{\lambda(t, s)}M(t, s)ds.
\end{equation}
The total charge $Q(t)$ of the system is given by:
\begin{equation*}
Q(t) := 4\pi q\int_{0}^{+\infty}s^{2}e^{\lambda(t,
s)}\int_{\mathbb{R}^{3}}f(t, s, v)dvds. \tag{5.10'}
\end{equation*}
Since the distribution function $f$ satisfies the Vlasov equation
one obtains the following conservation law (see Proposition 3.1,
\cite{noundjeu2}):
\begin{equation*}
\frac{\partial}{\partial t}\left( e^{\lambda}\int_{\mathbb{R}^{3}}
fdv \right) + \underset{\tilde{x}}{\div}\left(
e^{\mu}\int_{\mathbb{R}^{3}}\frac{v}{\sqrt{1 + v^{2}}}fdv \right)
= 0.
\end{equation*}
With the help of this, we obtain:
\begin{align*}
\dot{Q}(t) &= 4\pi q\int_{0}^{+
\infty}s^{2}\frac{\partial}{\partial t}\left(
e^{\lambda}\int_{\mathbb{R}^{3}} fdv \right)ds\\
&= 4\pi q\lim_{r \to
+\infty}\int_{0}^{r}s^{2}\frac{\partial}{\partial t}\left(
e^{\lambda}\int_{\mathbb{R}^{3}} fdv \right)ds\\
&= q\lim_{r \to + \infty}\int_{\mid y \mid \leq
r}\frac{\partial}{\partial t}\left(
e^{\lambda}\int_{\mathbb{R}^{3}} fdv \right)dy\\
&= -q\lim_{r \to + \infty}\int_{\mid y \mid \leq
r}\underset{y}{\div}\left(
e^{\mu}\int_{\mathbb{R}^{3}}\frac{v}{\sqrt{1 + v^{2}}}fdv
\right)dy\\
&= -q\lim_{r \to + \infty}\int_{\mid y \mid =
r}e^{\mu}\int_{\mathbb{R}^{3}}\frac{y . v}{r\sqrt{1 +
v^{2}}}fdvd\omega(y)\\
&= -4\pi q\lim_{r \to + \infty}re^{\mu(t, r)}N(t, r), \quad
\text{since} \, \int_{\mid y \mid = r}d\omega(y) = 4\pi r^{2}\\
&= 0, \quad \text{since} N(t) \, \text{is compactly supported}.
\end{align*}
Thus $Q(t)$ is a constant and we can write: $Q(t) = Q(0) := Q$. So
with the above we can deduce a bound of $e(t, r)$, since $\lambda
\geq 0$:
\begin{equation} \label{eq:5.11}
\mid e(t, r) \mid \leq Q/(4\pi r^{2}) \leq C\varepsilon^{-2} \leq
C.
\end{equation}
and we have a bound of the fourth term in the right hand side of
(\ref{eq:5.8}), since $\lambda + \mu \leq 0$:
\begin{equation} \label{eq:5.12}
0 \leq \mid qe^{\mu + \lambda}e(t, r) \mid \leq C.
\end{equation}
Using the above inequalities, we obtain the following estimate for
$\dot{w}$:
\begin{equation} \label{eq:5.13}
-C\sqrt{C + w^{2}} + 4\pi re^{\mu + \lambda}(wk - p\sqrt{1 +
u^{2}}) \leq \dot{w} \leq C + 4\pi re^{\mu + \lambda}(wk -
p\sqrt{1 + u^{2}})
\end{equation}
Since $0 \leq 4\pi re^{\mu + \lambda} \leq C$ on the support of
$f$, we concentrate on the quantity $wk - p\sqrt{1 + u^{2}}$. Set
\begin{equation*}
\tilde{w} := r^{-1}\tilde{x} . \tilde{v}, \quad \tilde{u} := \mid
\tilde{v} \mid, \quad \tilde{L} := r^{2}\tilde{u}^{2} - (\tilde{x}
. \tilde{v})^{2}.
\end{equation*}
Then
\begin{equation} \label{eq:5.14}
\begin{aligned}
wk - p\sqrt{1 + u^{2}} &= \int_{\mathbb{R}^{3}}f(t, \tilde{x},
\tilde{v})\left( w\tilde{w} - \sqrt{1 +
u^{2}}\frac{\tilde{w}^{2}}{\sqrt{1 + \tilde{u}^{2}}}
\right)d\tilde{v}\\
& \qquad + \frac{1}{2}\sqrt{1 + u^{2}}e^{2\lambda}e^{2}.
\end{aligned}
\end{equation}
Using once again (\ref{eq:5.10}), and the fact that the total
charge is conserved with time $t$ one obtains:
\begin{equation*}
0 \leq e^{2\lambda}e^{2} \leq C/r^{4} \leq C\varepsilon^{-4} \leq
C.
\end{equation*}
So, the above and (\ref{eq:5.6}), yield:
\begin{equation*}
0 \leq 1/2\sqrt{1 + u^{2}}e^{2\lambda}e^{2} \leq C\sqrt{C + w^{2}}
\leq C\sqrt{C + P_{s}^{2}(t)}.
\end{equation*}
Next, for $w > 0$, if $\tilde{w} < 0$ then the first term in the
right hand side of (\ref{eq:5.14}) is negative and we obtain the
following estimate for $\dot{w}$:
\begin{equation} \label{eq:5.15}
\dot{w} \leq C + C\sqrt{C + P_{s}^{2}(t)}.
\end{equation}
For $\tilde{w} > 0$, one has (see \cite{rein2}):
\begin{equation*}
\int_{\mathbb{R}^{3}}f\left( w\tilde{w} - \sqrt{1 +
u^{2}}\frac{\tilde{w}^{2}}{\sqrt{1 + \tilde{u}^{2}}}
\right)d\tilde{v} \leq CP_{s}(t)\ln (1 + P_{s}^{2}(t))
\end{equation*}
and using (\ref{eq:5.14}) we deduce:
\begin{equation*}
\dot{w} \leq C + CP_{s}(t) + CP_{s}(t)\ln (1 + P_{s}^{2}(t)).
\end{equation*}
Set $\varphi(x) = \ln (1 + x^{2}) - 1$, $\alpha = \sqrt{e - 1}$.
Then $\varphi(\alpha) = 0$, $\varphi(x) > 0$ for $x \in ]\alpha, +
\infty[$ and $\varphi(x) < 0$ for $x \in [0, \alpha[$. Suppose
that $P_{s}(t) > \alpha$ for $t \in [0, T[$. Then $P_{s}(t) \leq
P_{s}(t)\ln(1 + P_{s}^{2}(t))$ and we deduce from the inequality
above that:
\begin{equation} \label{eq:5.16}
\dot{w} \leq C + CP_{s}(t)\ln(1 + P_{s}^{2}(t)) + C\frac{\ln(1 +
P_{s}^{2}(t))}{P_{s}(t)}.
\end{equation}
If $0 < P_{s}(t) < \alpha$ for $t \in [0, T[$, then $P_{s}(t)\ln(1
+ P_{s}^{2}(t)) \leq P_{s}(t)$, and we deduce from the above that:
\begin{equation} \label{eq:5.17}
\dot{w} \leq C + CP_{s}(t).
\end{equation}
Let $w(\tau)$ be the values of $w$ along a characteristic and set:
\begin{equation*}
t_{0} := \inf \{ \tau \geq 0: w(s) \geq 0 \, \text{for} \, s \in
]\tau, t( \}.
\end{equation*}
Then $w(t_{0}) \leq C$ and (\ref{eq:5.16}) yields by integration:
\begin{equation*}
w(t) \leq C + C\int_{t_{0}}^{t}\left( P_{s}(\tau)\ln(1 +
P_{s}^{2}(\tau)) + \frac{\ln(1 + P_{s}^{2}(\tau))}{P_{s}(\tau)}
\right)d\tau.
\end{equation*}
Set $\bar{P_{s}}(\tau) := \max (0, P_{s}(\tau))$, then for $t_{0}
= 0$, we can write:
\begin{equation*}
\bar{P_{s}}(t) \leq C + C\int_{0}^{t}\left( \bar{P_{s}}(\tau)\ln(1
+ \bar{P_{s}}^{2}(\tau)) + \frac{\ln(1 +
\bar{P_{s}}^{2}(\tau))}{\bar{P_{s}}(\tau)} \right)d\tau
\end{equation*}
and using the generalized form of the Gronwall inequality (see
\cite{taylor}, page 26), one has:
\begin{equation*}
\bar{P_{s}}(t) \leq z(t),
\end{equation*}
where $z$ is a solution of the following differential equation:
\begin{equation*}
\dot{z}(t) = Cz(t)\ln(1 + z^{2}(t)) + C\frac{\ln(1 +
z^{2}(t))}{z(t)}
\end{equation*}
and since $z(t) = \sqrt{\text{exp}(e^{2Ct}) - 1}$ is a solution of
the equation above, one finally obtains:
\begin{equation} \label{eq:5.18}
P_{s}(t) \leq \bar{P_{s}}(t) \leq \sqrt{\text{exp}(e^{2Ct}) - 1}
\leq C.
\end{equation}
Also, we integrate (\ref{eq:5.17}) on $[0, t]$ and obtain:
\begin{equation*}
\bar{P_{s}}(t) \leq C + C\int_{0}^{t}\bar{P_{s}}(\tau)d\tau
\end{equation*}
and deduce by Gronwall's lemma the following bound of $P_{s}(t)$:
\begin{equation*}
P_{s}(t) \leq \max(\alpha, e^{Ct}), \quad t \in [0, T[.
\end{equation*}
Note that the above is valid in the case $P_{s}(t) < 0$. We now
look for a bound of $P_{i}(t)$. Suppose $P_{i}(t) < 0$ and
consider $w$ in $\supp f$ with $P_{i}(t) < w < 0$. For $\tilde{w}
\leq 0$, we get (see \cite{rein2}):
\begin{equation*}
w\tilde{w} - \sqrt{1 + u^{2}}\frac{\tilde{w}^{2}}{\sqrt{1 +
\tilde{u}^{2}}} \geq -C\frac{\mid \tilde{w} \mid}{\sqrt{1 +
\tilde{w}^{2}}}
\end{equation*}
and one uses (\ref{eq:5.18}) to obtain, for $w < 0 < \tilde{w}
\leq P_{s}(t)$:
\begin{equation*}
w\tilde{w} - \sqrt{1 + u^{2}}\frac{\tilde{w}^{2}}{\sqrt{1 +
\tilde{u}^{2}}} \geq Cw - C\sqrt{C + w^{2}}.
\end{equation*}
So, we deduce:
\begin{equation*}
\int_{\mathbb{R}^{3}}f\left( w\tilde{w} - \sqrt{1 +
u^{2}}\frac{\tilde{w}^{2}}{\sqrt{1 + \tilde{u}^{2}}}
\right)d\tilde{v} \geq -CP_{i}^{2}(t)\frac{1}{\sqrt{1 + w^{2}}} +
C\bar{P_{s}}(t)(w - \sqrt{C + w^{2}}).
\end{equation*}
\begin{equation*}
\dot{w} \geq -C\sqrt{C + w^{2}} - CP_{i}^{2}(t)\frac{1}{\sqrt{1 +
w^{2}}} + Cw
\end{equation*}
\begin{equation*}
\dot{w}^{2} = 2w\dot{w} \leq C + CP_{i}^{2}(t).
\end{equation*}
Set
\begin{equation*}
t_{1} := \inf \{ \tau \geq 0: \, w(s) \leq 0 \quad \text{for}
\quad s \in ]\tau, t[ \}.
\end{equation*}
Then $0 \geq w(t_{1}) \geq -C$ and one proceeds as before to
obtain:
\begin{equation*}
P_{i}^{2}(t) \leq C + C\int_{0}^{t}P_{i}^{2}(\tau)d\tau, \quad
P_{i}(t) < 0,
\end{equation*}
and the Gronwall lemma yields:
\begin{equation*}
P_{i}^{2}(t) \leq Ce^{Ct} \leq C, \, \text{for} \, t \in [0, T[.
\end{equation*}
Besides, if $P_{i}(t) \geq 0$ then
\begin{equation*}
0 \leq P_{i}(t) \leq P_{s}(t) \leq C,
\end{equation*}
and $P_{i}(t)$ is also bounded for that case. So, $P(t)$ is
bounded and Theorem 5.1 is proved.

We now focus on what happens when we are far from the center.
\section{The Exterior Problem}
Let $r_{1}$ and $T$ be positive real numbers. Consider the
exterior region:
\begin{equation*}
W(T, r_{1}) := \{ (t, r): \, 0 \leq t < T, \quad r \geq r_{1} +
t\}.
\end{equation*}
In what follows, the initial value problem (\ref{eq:2.2}),
(\ref{eq:2.3}), (\ref{eq:2.4}), (\ref{eq:2.5}), (\ref{eq:2.9}),
and (\ref{eq:2.10}) are going to be studied on a domain of this
kind. Take $\overset{\circ}{f} \geq 0$, spherically symmetric,
$C^{\infty}$, compactly supported and defined on the region $\mid
\tilde{x} \mid \geq r_{1}$ and let $(\overset{\circ}{\lambda},
\overset{\circ}{\mu}, \overset{\circ}{e})$ be a regular solution
of the constraint equations on that region. Since we are away from
the center of symmetry, we have to change the boundary conditions
(\ref{eq:2.9}) as follows: Let $m_{\infty} \geq 4\pi
\int_{r_{1}}^{\infty}s^{2}\rho(0, s)ds$, and $M_{\infty} \geq
\int_{r_{1}}^{\infty}s^{2}e^{\overset{\circ}{\lambda}(s)}M(0,
s)ds$. Take any solution of (\ref{eq:2.2}), (\ref{eq:2.3}),
(\ref{eq:2.4}) and (\ref{eq:2.5}) on $W(T, r_{1})$ and define
\begin{equation} \label{eq:6.1}
\begin{cases}
m(t, r) := m_{\infty} - 4\pi \int_{r}^{\infty}s^{2}\rho(t, s)ds\\
\bar{M}(t, r) := M_{\infty} - \int_{r}^{\infty}s^{2}e^{\lambda(t,
s)}M(t, s)ds
\end{cases}
\end{equation}
Then for all $r \geq r_{1}$, $m(0, r) \geq 0$. So, we replace
(\ref{eq:2.9}) by:
\begin{equation} \label{eq:6.2}
\begin{cases}
e^{-2\lambda(t, r)} = 1 - 2\frac{m(t, r)}{r}; \quad \underset{r
\rightarrow \infty}{\lim} \lambda(t, r) =
\underset{r \rightarrow \infty}{\lim}\mu(t, r) = 0\\
e(t, r) = \frac{q}{r^{2}}e^{-\lambda(t, r)}\bar{M}(t, r); \quad
\underset{r \rightarrow \infty}{\lim}e(t, r) = 0
\end{cases}
\end{equation}
Note that the restriction of a solution of the original problem
with boundary conditions (\ref{eq:2.9}) on $W(T, r_{1})$,
satisfies (\ref{eq:6.2}) if $m_{\infty}$ is chosen to be the A.D.M
mass of the solution on the entire space, and $M_{\infty} =
Q/(4\pi q)$, $Q$ being the total charge given by (5.10'). Besides,
to obtain a local existence theorem as in \cite{noundjeu2}, we use
the fact that (\ref{eq:6.2}) must hold on the initial hypersurface
$t = 0$ that leads to the following condition:
\begin{equation} \label{eq:6.3}
m_{\infty} - \underset{\mid \tilde{x} \mid \geq r}{\int}\left(
\int_{\mathbb{R}^{3}}\sqrt{1 + v^{2}}\overset{\circ}{f}(\tilde{x},
v)dv \right)d\tilde{x} < \frac{r}{2}, \quad r \geq r_{1}
\end{equation}
\begin{theorem} [Local existence and uniqueness] \label{T:6.1}
Fix $m_{\infty} > 0$. Let \\ $\overset{\circ}{f} \geq 0$ be a
spherically symmetric function on the region $\mid \tilde{x} \mid
\geq r_{1}$ which is $C^{\infty}$ and compactly supported. Let
$(\overset{\circ}{\lambda}, \overset{\circ}{\mu},
\overset{\circ}{e})$ be a regular solution of the constraint
equations on the region $r \geq r_{1}$. Suppose that
(\ref{eq:6.3}) holds and that
\begin{equation} \label{eq:6.4}
\underset{\mid \tilde{x} \mid \geq r_{1}}{\int}\left(
\int_{\mathbb{R}^{3}}\sqrt{1 + v^{2}}\overset{\circ}{f}(\tilde{x},
v)dv \right)d\tilde{x} < m_{\infty}
\end{equation}
\begin{equation*}
\underset{\mid \tilde{x} \mid \geq r_{1}}{\int}
e^{\overset{\circ}{\lambda}(\tilde{x})}\left(
\int_{\mathbb{R}^{3}}\overset{\circ}{f}(\tilde{x}, v)dv
\right)d\tilde{x} \leq 4 \pi M_{\infty}. \tag{6.4'}
\end{equation*}
Then there exists a unique regular solution of \ref{eq:2.2}),
(\ref{eq:2.3}), (\ref{eq:2.4}) and (\ref{eq:2.5}) on the region
$W(T, r_{1})$ with $(\overset{\circ}{f}, \overset{\circ}{\lambda},
\overset{\circ}{\mu}, \overset{\circ}{e})$ which satisfies the
boundary conditions (\ref{eq:6.2}).
\end{theorem}
\textbf{Proof:} Let $\overset{\circ}{f}$ be as in Theorem 6.1.
Suppose
\begin{equation*}
\supp \overset{\circ}{f} \subset B(R) \times B(R'), \quad R, R' >
0
\end{equation*}
where $B(R)$ is the open ball of $\mathbb{R}^{3}$ with radius $R$.
We only have to check that one can construct a solution satisfying
the constraints and use \cite{noundjeu2} to obtain the desired
result. If $r_{1} \geq R$, then the constraint equations reduces
to the static Einstein-Maxwell system and a solution for this
system on the region $r \geq r_{1}$ is a part of the
Reissner-Nordstr\"om solution \cite{hawking}. Now, if $r_{1} < R$,
since the constraint equations is invariant with translation we
can use perturbation techniques to establish a solution for the
corresponding Cauchy problem provided $\overset{\circ}{f}$
satisfies an appropriate condition like (\ref{eq:3.1}). The reader
can refer to \cite{noundjeu1}, to have more details. So Theorem
6.1 is proved.

We now derive a corresponding continuation criterion, with proof
being as that of Theorem 3.2 in \cite{rein2}. Before doing so, we
mean by maximal interval of existence for the exterior problem,
the largest region $W(T, r_{1})$ on which a solution exists with
given initial data and the parameters $r_{1}$ and $m_{\infty}$.
\begin{theorem}[Continuation Criterion] \label{T:4.2}
Fix $m_{\infty} > 0$. Let $\overset{\circ}{f} \geq 0$ be a
spherically symmetric function on the region $\mid \tilde{x} \mid
\geq r_{1}$ which is $C^{\infty}$, compactly supported and
satisfies (\ref{eq:6.3}) and (\ref{eq:6.4}). Let
$(\overset{\circ}{\lambda}, \overset{\circ}{\mu},
\overset{\circ}{e})$ be a regular solution of the constraints on
the region $r \geq r_{1}$ so that (6.4') holds. Let $(f, \lambda,
\mu, e)$ be a regular solution of (\ref{eq:2.2}), (\ref{eq:2.3}),
(\ref{eq:2.4}) and (\ref{eq:2.5}) on $W(T, r_{1})$ with initial
data $(\overset{\circ}{f}, \overset{\circ}{\lambda},
\overset{\circ}{\mu}, \overset{\circ}{e})$. If $T < \infty$ and
$W(T, r_{1})$ is the maximal interval of existence, then $P$ is
bounded.
\end{theorem}
\section{The Regularity Theorem}
Here we consider the initial boundary value problem
(\ref{eq:2.2}), (\ref{eq:2.3}), (\ref{eq:2.4}), (\ref{eq:2.5}),
(\ref{eq:2.9}) and (\ref{eq:2.10}). The following result shows
that if a solution of this Cauchy problem develops a singularity
at all, the first one must be at the center.
\begin{theorem} \label{T:7.1}
Let $(\overset{\circ}{\lambda}, \overset{\circ}{\mu},
\overset{\circ}{e})$  be a solution of the constraint equations.
Let $(f, \lambda, \mu, e)$ be the corresponding regular solution
defined on a time interval $[0, T[$. Suppose that there exists an
open neighborhood $U$ of the point $(T, 0)$ such that
\begin{equation} \label{eq:7.1}
\sup \{ \mid v \mid: (t, \tilde{x}, v) \in \supp f \cap (U \times
\mathbb{R}^{3}) \} < \infty.
\end{equation}
Then $(f, \lambda, \mu, e)$ extends to a regular solution on $[0,
T'[$ for some $T' > T$.
\end{theorem}
\textbf{Proof:} See \cite{rein2}.
\\
\\
\textbf{Acknowledgment:} The author is grateful to A.D. Rendall
for helpful suggestions. This work was supported by a research
grant from the VolkswagenStiftung, Federal Republic of Germany.

\end{document}